\documentclass{iaus}
\usepackage{graphicx}

\title[Variable Stars in Tucana and LGS3] 
{The ACS LCID Project: \break Variable Stars in Tucana and LGS3}

\author[Edouard J. Bernard for the LCID Team]   
{Edouard J. Bernard for the LCID Team\thanks{Local Cosmology from Isolated Dwarfs:
http://www.iac.es/project/LCID}}

\affiliation{Instituto de Astrof\'isica de Canarias, E-38205 La Laguna, Tenerife, Spain
             \break email: ebernard@iac.es}

\pubyear{2007}
\volume{241}  
\pagerange{1--2}
\date{?? and in revised form ??}
\jname{Stellar Populations as Building Blocks of Galaxies}
\editors{A. Vazdekis \& R. Peletier, eds.}
\begin{document}

\maketitle

\begin{abstract}
 We present preliminary results concerning the search for short-period
 variable stars in Tucana and LGS3 based on very deep HST/ACS imaging. 
 In the one chip per galaxy we studied so far, a total of
 230 and 80 candidates variables were found, respectively. 
 For Tucana, we identified 134 of them as RR Lyrae stars (RRL) pulsating in the
 fundamental mode (RR$ab$), 51 in the first-overtone mode (RR$c$), and 37 in both
 modes simultaneoulsy (RR$d$), as well as four candidate anomalous Cepheids (AC).
 In the case of LGS3, we found 45 RR$ab$ and 5 RR$c$, plus three candidates RR$d$
 and five candidate AC. The metallicities obtained from the mean period of the
 RR$ab$ are [Fe/H]$_{Tuc}=-1.7$ and [Fe/H]$_{LGS3}=-1.8$.
\end{abstract}

\firstsection 

\section{Introduction}

 Pulsating variable stars play a fundamental role in the study of stellar
 populations and in Cosmology, as their pulsational properties are
 traditionally used to determine distances and metallicities, and put
 constraints on stellar physical properties. Because the pulsations occur
 at a particular phase of their evolution depending on the star mass, these
 stars trace the spatial distribution of stellar populations of given ages, 
 therefore highlighting the eventual radial trends across the studied galaxy
 (e.g., Phoenix: \cite{gal04}, Leo I: \cite{bal04}). This, in turn, hints on
 the star formation history and formation mechanisms of the host galaxy.
 
 In a cycle 14 HST/ACS program we obtained, for the first time, very deep
 ($V\sim29$) multi-epoch images of five {\it isolated} dwarf galaxies of the
 Local Group: the dwarf spheroidals (dSph) Tucana and Cetus, the dwarf irregulars
 (dIrr) IC1613 and Leo A, and the so-called transition type dIrr/dSph LGS3. See
 Gallart \etal\ (these proceedings) for an overview of the LCID project, and
 Monelli \etal\ (these proceedings) for a description of the data and data reduction.

 \section{Variable stars search and first results}\label{sec:results}
 
 The DAOPHOT/ALLFRAME suite of programs (\cite{ste94})
 was used to obtain the instrumental photometry of the stars on the individual
 images. The candidate variables were extracted from the star list using the
 variability index given by DAOMASTER. Figure 1 ({\it left \& middle panels})
 shows the candidate variables
 in each galaxy, highlighted on a portion of the CMDs centered on the instability
 strip. Period search for the candidates was done using an implementation of the 
 phase-dispersion minimization method (\cite{ste78}) taking into account the
 information from both bands simultaneously.

 \begin{figure}
\centerline{\scalebox{0.49}{\includegraphics{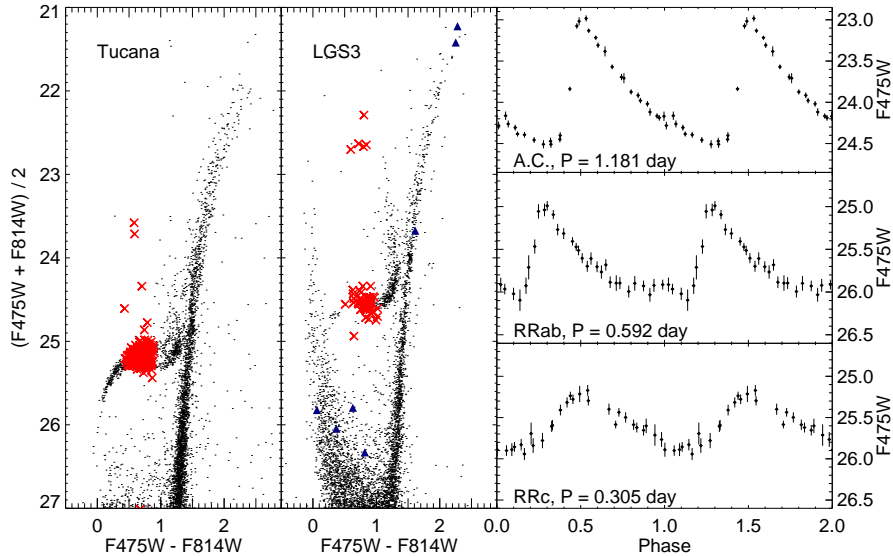}}}
  \caption{Color-magnitude diagrams of Tucana ({\it left}) and LGS3
  ({\it middle}). Variables for which a period was found are shown as red crosses.
  Candidate variables are indicated by blue triangles. {\it Right:} Light-curves
  for some variables in Tucana. 
  }\label{fig:cmd}
\end{figure}

 We first focused on the candidate variables found on only
 one chip of the ACS. For Tucana, we found 134 RR$ab$
 and 51 RR$c$ with mean periods of 0.601 and 0.350 days, respectively, as
 well as 37 RR$d$. This large percentage of RR$d$ ($\sim$17\%) is similar
 to that found by \cite{cle06} in the Fornax dSph. The four variables brighter
 than the HB are most likely AC. Typical light-curves are shown in 
 Fig.~\ref{fig:cmd} ({\it right panel}).

 For LGS3, the small number of datapoints (12 in each band, vs. 32 for Tucana)
 made uncertain the period estimates,
 and the particular temporal sampling created strong aliasing. However, the
 period-amplitude diagram supported our
 choice of the period. It also showed that the amplitude of the RRL stars is
 systematically smaller in LGS3 than in Tucana. Although some of the lowest
 amplitudes are due to the lack of observations at maximum light, this trend
 has been noted for dSph and globular clusters having a very red horizontal
 branch (HB) (\cite{pri05}, and references therein). Indeed, the HB of LGS3
 is mainly red, with RRL stars 
 located near the red edge of the instability strip.

 From the mean period of the RR$ab$, we calculated the metallicity of the old
 population using the relation of \cite{sar06}. We obtained [Fe/H]$_{Tuc}=-1.7$
 and [Fe/H]$_{LGS3}=-1.8$, which is consistent with the values found through
 isochrone fitting.

\begin{acknowledgments}
This project is partially supported by a Marie Curie fellowship (MEST-CT2004-504604),
the Spanish Ministry of Education \& Science (AYA2004-06343) and the IAC (P3/94).
\end{acknowledgments}

\end{document}